\documentclass[preprint2]{aastex6}

\slugcomment{Draft version 2 -- \today}

\shorttitle{Turbulent drift reduction}
\shortauthors{Engelbrecht et al.}

\begin{document}


\title{Towards a greater understanding of the reduction of drift coefficients in the presence of turbulence}


\author{N.E. Engelbrecht\altaffilmark{1}, R.D. Strauss \altaffilmark{1,2}, J.A. le Roux\altaffilmark{3,4} and R.A. Burger \altaffilmark{1}}
\email{n.eugene.engelbrecht@gmail.com}


\altaffiltext{1}{Center for Space Research, North-West University, Potchefstroom, 2522, South Africa}
\altaffiltext{2}{National Institute for Theoretical Physics (NITheP), Gauteng, South Africa}
\altaffiltext{3}{Center for Space Plasma and Aeronomic Research, University of Alabama in Huntsville, Huntsville, AL 3585, USA}
\altaffiltext{4}{Department of Space Science, University of Alabama in Huntsville, Huntsville, AL 35899, USA}


\begin{abstract}

Drift effects play a significant role in the transport of charged particles in the heliosphere. A turbulent magnetic field is also known to reduce the effects of particle drifts. The exact nature of this reduction, however, is not clear. This study aims to provide some insight into this reduction, and proposes a relatively simple, tractable means of modelling it that provides results in reasonable agreement with numerical simulations of the drift coefficient in a turbulent magnetic field. 

\end{abstract}





\section{Introduction}
\label{sec-1}

Drift due to magnetic field gradients and curvatures play a central role in the transport of charged particles in a plasma. Cosmic rays in the heliosphere experience drifts, not only due to the gradient and curvature of the heliospheric magnetic field, but also due to the heliospheric current sheet, a surface over which the sign of the heliospheric magnetic field is reversed. These drifts have long been known to have significant effects on cosmic-ray transport, and hence on cosmic-ray modulation \citep[e.g][]{jl1977,Jokipii1977,Jokipii1979,jt1981}, even in the heliosheath \citep{kota}. Drift effects account for the 22-year cycle observed in cosmic-ray intensities \citep{jt1981}, lead to a strong dependence of observed cosmic-ray intensities on the solar tilt angle \citep{lockwoodwebber2005} and heliospheric magnetic field polarity \citep{webberlockwood2005}, as well as having a significant influence on observed global cosmic-ray modulation phenomena such as observed latitude gradients \citep{heber1996,zhang1997,desimone2011}. Drift effects may even be of importance to the study of solar energetic particles \citep[e.g.][]{dalla2013}. The drift coefficient, which enters the \citet{parker1965} cosmic-ray transport equation via the off-diagonal elements of the diffusion tensor, can, in the weak scattering limit, be expressed by \citep[e.g.][]{formanetal1974}
\begin{equation}
\kappa_{A}^{ws}=\frac{v}{3}R_{L},
\label{wska}
\end{equation}
with $R_{L}$ the maximal gyroradius and $v$ the particle speed. Particle drift coefficients have been shown theoretically and by means of numerical test-particle simulations \cite[see, e.g.,][]{b90,jokipii1993,fs95,giacaloneetal1999,Candia2004,Minnie_etal2007b,ts2012} to be reduced in the presence of turbulence. It is interesting to note that \citet{park65} incorporated a reduction factor in the off-diagonal elements of the diffusion tensor, albeit due to isotropic scattering. Given the importance of drift in any study of cosmic-ray modulation, this reduction needs to be carefully modelled, as numerical cosmic-ray modulation studies also indicate that better agreement of model results with spacecraft observations can be found if the cosmic-ray drift coefficient at low to intermediate values were smaller than the weak-scattering value of Eq. \ref{wska} \citep[e.g.][]{pb90}, and are very sensitive to the choice made as to the drift-reduction factor \citep{EB2015,np2015}. Furthermore, such modulation studies have shown a marked solar-cycle dependence of the factor by which the weak-scattering drift coefficient needs to be reduced, so as to fit spacecraft observations of cosmic ray intensities \citep[e.g.][]{Ndii,rex}. Cosmic ray modulation studies have long employed an \textit{ad hoc} form for the reduced drift coefficient \citep[see, e.g.,][]{burgeretal2000,ep2016,rendani}, given by
\begin{equation}
\kappa_{A}=\frac{\beta P}{3B_{0}}\frac{(P/P_0)^2}{1+(P/P_0)^2},
\label{eq:dofka}
\end{equation}
with $B_{0}$ the background magnetic field magnitude, $\beta$ the ratio of the particle's speed to that of light, $P$ the particle rigidity, and $P_{0}$ an \textit{ad hoc} parameter, in units of GV, that is chosen so as to achieve model agreement with a particular spacecraft dataset. Different values for $P_0$, however are required to fit different sets of spacecraft data \citep{EB2015}, and it must be noted that, in such modulation studies, a large perpendicular coefficient would act so as to mask the effects of drift, even for large values of the drift coefficient \cite[see, e.g.,][]{k89,kota}.

Numerical test particle simulations, where the Newton-Lorentz equation is solved for an ensemble of test particles in various pre-specified turbulent magnetic field conditions, do reveal some details as to the exact nature of the reduction of the drift coefficient. \citet{giacaloneetal1999} first showed, by means of such simulations for simulated composite slab/2D turbulence \citep[see, e.g.][]{bieberetal1994} and isotropic turbulence, that drift coefficients are indeed reduced under such circumstances, a result confirmed for isotropic turbulence by \citet{Candia2004} and for composite slab/2D turbulence by \citet{Minnie_etal2007b}. \citet{Minnie_etal2007b} studied this effect for both a uniform background magnetic field as well as a background field with an imposed spatial gradient, finding the same levels of reduction for the drift coefficient in each case when the same turbulence conditions are used. They also showed that the total drift motion of the particle is not completely described by the off-diagonal elements of the diffusion tensor, and that, due to the scattering of particles, a proper understanding of the drifts of these particles requires an understanding of the symmetric elements of the diffusion tensor, \textit{i.e.} the parts that govern diffusion parallel and perpendicular to the background field. Furthermore, for their simulations incorporating a background field with a gradient, these authors also report a reduction in the drift velocity of particles in the presence of turbulence, as would be expected from the behaviour of the corresponding drift coefficient, which \citet{Minnie_etal2007b} found to agree with that calculated from their simulations performed assuming a uniform background magnetic field.

\citet{ts2012} performed extensive simulations of the drift coefficient, for different turbulent geometries, and different wavenumber-dependencies of the energy-containing range on the assumed turbulence power spectral form. In line with the previously mentioned studies, \citet{ts2012} find, for isotropic and composite turbulence, that the drift coefficient is essentially the weak scattering coefficient given in Eq. \ref{wska} for very low levels of turbulence, becoming ever more reduced as turbulence levels increase, with the amount of reduction decreasing for a given turbulence level as particle energy is increased. Interestingly, \citet{ts2012} show that, no matter the strength, pure slab turbulence simply does not reduce the computed drift coefficient from the weak scattering value. These authors also report a relatively weak dependence of the drift-reduction factor on particle rigidity and on the energy-range spectral index of the 2D fluctuation spectrum. It should be noted that all the abovementioned simulations were performed assuming axisymmetric, transverse magnetostatic turbulent fluctuations, and also that, although the simulations of \citet{ts2012} agree qualitatively (where comparable) with the results of the studies of \citet{Candia2004} and \citet{Minnie_etal2007b}, they do not agree quantitatively.

Due to the extreme complexity of a self-consistent theoretical approach to the reduction of drift effects in the presence of turbulence \citep[see, e.g.,][]{Stawicki2005,leroux2007}, there have been relatively few attempts at theoretical treatments of this problem. Numerical studies of the drift coefficient report on fits to the turbulence-reduced drift coefficient \citep{Candia2004,ts2012}, but these are potentially of limited use to, e.g., modulation studies, as the simulated turbulence conditions assumed in these studies may not necessarily be representative of heliospheric conditions. An example of such a fit is presented by \citet{ts2012}, where
\begin{equation}
\kappa_{A}=\frac{v}{3}R_{L}\frac{1}{1+a(\delta B_{T}^{2}/B^{2}_{0})^{d}},
\label{eq:tsdrift}
\end{equation}
with $a$ and $d$ fitting constants that change with different turbulence geometries assumed in the simulations, and $\delta B_{T}^{2}$ the (total) magnetic variance. Note that this expression is similar to what was suggested by \citet{jk1989}. \citet{bm1997}, considering the effects of transverse turbulent fluctuations on the unperturbed particle orbits, find that the drift coefficient is given by
\begin{equation}
\kappa_{A}=\frac{vR_L}{3}\frac{(\Omega \tau)^2}{1+(\Omega \tau)^2} \label{aap0}
\end{equation}
where $\Omega$ is the (unperturbed) particle gyrofrequency, and $\tau$ some decorrelation time. The product of these two quantities they model using $\Omega \tau=2R_{L}/3D_{\perp}$, where $R_{L}$ is the maximal (unperturbed) particle Larmor radius, and $D_{\perp}$ the field line random walk (FLRW) diffusion coefficient \citep[see, e.g.,][]{matthaeusetal1995}, given for slab/2D composite turbulence by
\begin{equation}
D_{\perp}=\frac{1}{2}\left(D_{sl}+\sqrt{D_{sl}^{2}+4D_{2D}^{2}}\right).
\label{eq:dperp}
\end{equation}
where
\begin{equation}
D_{sl}=\frac{1}{2}\frac{\delta B_{s}^{2}}{B_{o}^{2}}\lambda_{c,s},
\label{eq:dslabm}
\end{equation}
and
\begin{equation}
D_{\perp}=\frac{\sqrt{\delta B_{2D}^{2}/2}}{B_{o}}\lambda_{u},
\label{eq:d2dm}
\end{equation}
with $\delta B_{s}^{2}$ and $\delta B_{2D}^{2}$ variances, respectively, $\lambda_{c,s}$ the slab correlation scale, and $\lambda_{u}$ the 2D ultrascale \cite[see, e.g.,][]{2007_Mattheaus_etal_ApJ}. Although providing a tractable expression for the turbulence-reduced drift coefficient, \citet{Burger2010} showed that the \citet{bm1997} drift-reduction factor simply does not fit the simulation results of \citet{Minnie_etal2007b}, whether they pertained to the drift coefficient or the drift velocity. These authors went on to propose another form for $\Omega \tau$, such that
\begin{equation}
\Omega\tau=\frac{11}{3}\frac{\sqrt{R_{L}/\lambda_{c}}}{(D_{\perp}/\lambda_{c})^{g}},
\label{eq:BVomtau}
\end{equation}
where $g=0.3\log(R_{L}/\lambda_{c})+1.0$ and $\lambda_{c}$ is the slab correlation length. This form then fit the \citet{Minnie_etal2007b} simulations very well, and has been used with some success in cosmic ray modulation studies \citep[e.g.][]{EB2013}, but the generality of this result is questionable, as it remains to be seen whether this highly parametrized fit would also agree with the results of simulations performed assuming turbulence conditions very different to those assumed by \citet{Minnie_etal2007b}. Lastly, only the complicated results presented by \citet{Stawicki2005} and \citet{leroux2007} predict the lack of drift reduction seen in the simulation results of \citet{ts2012} for pure magnetostatic slab turbulence.

The question that this study attempts to answer, then, is whether one can derive a simple, tractable expression for the drift-reduction factor that is in agreement with what is known of this quantity from numerical simulations, and which in principle can be applied in the broad range of turbulence conditions typically encountered by, e.g., galactic cosmic rays and solar energetic particles as they traverse the heliosphere. Firstly, from a simplistic analysis of the drift velocity of a charged particle in a turbulent magnetic field we show that one can readily derive an expression for the drift reduction factor similar to the fits proposed by \citet{ts2012} that yields results that, in limiting cases, bound the simulation results of e.g. \citet{Minnie_etal2007b}. In Section \ref{sec-bm} a new drift-reduction factor is derived, broadly following the approach taken by \citet{bm1997}, which not only produces results in reasonably good agreement with the simulations of \citet{Minnie_etal2007b} for both the drift velocity and drift coefficient, but also returns the weak-scattering drift coefficient should the assumption of magnetostatic, purely slab turbulence be made. The last section provides a discussion of the abovementioned results.

\section{A first-order approach to the effect of turbulence on cosmic ray drift coefficients}
\label{sec-2}

In general, the pitch-angle average guiding center drift velocity of a particle with momentum $p$ and charge $q$ in a fluctuating magnetic field $B$ is given by

\begin{equation}
\langle \vec{v}_{d} \rangle = \left\langle \frac{pv}{3q} \nabla \times \frac{\vec{B}}{B^2} \right\rangle, \label{aap1}
\end{equation}
with angle brackets denoting a suitable time average. Using a Reynold's decomposition of the magnetic field, the magnetic field can be written as the sum of a large scale $\vec{B_0}$ and fluctuating transverse $\vec{b}$ components such that
\begin{equation}
\vec{B} = \vec{B_0} + \vec{b}.
\end{equation}
Note that the assumption of transverse fluctuations is made throughout this study. The above, when substituted into Eq. \ref{aap1}, yields
\begin{eqnarray}
\langle \vec{v}_{d} \rangle &\approx& \frac{pv}{3q} \nabla \times \left\langle\frac{\vec{B}}{B^2} \right\rangle \nonumber \\
 &\approx& \frac{pv}{3q} \nabla \times \frac{\vec{B_0}}{B_0^2  + \langle   b^2\rangle} \nonumber \\
 &=&  \nabla \times \left( \frac{pv}{3qB_0}\right)  \left(\frac{\vec{B_0}}{B_0} \right) \left(\frac{B_0^2}{B_0^2  + \langle   b^2\rangle}\right).  \label{aaap2}
\end{eqnarray}
In the above equations it is assumed that the turbulence is weak (such that $b \ll B_0$) and vanishes when an appropriate long-term time-averaging is performed ($\langle   \vec{b}\rangle =0$). Moreover, due to the assumption of transverse turbulence,
\begin{equation}
B^2 = \langle \vec{B} \cdot \vec{B}\rangle = \langle\vec{B_0} \cdot \vec{B_0}\rangle + 2 \langle\vec{b} \cdot \vec{B_0}\rangle + \langle\vec{b} \cdot \vec{b}\rangle = B_0^2 + \langle b^2 \rangle.
\end{equation}

\begin{figure*}
    \centering
    \includegraphics[width=0.95\textwidth]{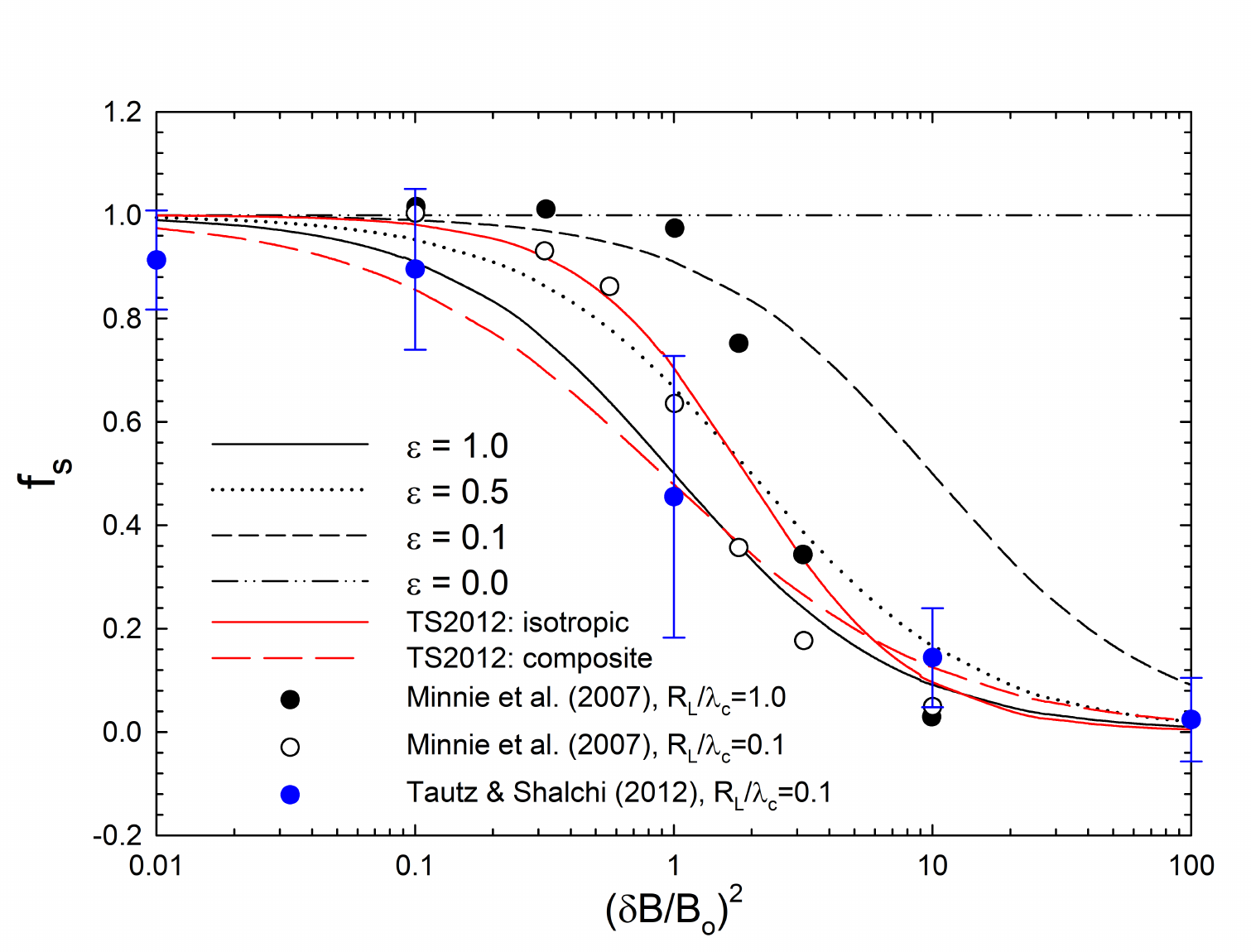}
    \caption{Drift reduction function (Eq. \ref{aap6}) for different choices of $\epsilon$ (black lines), with fits done by \citet{ts2012} to the drift reduction factor they find from their simulations of drift coefficients in the presence of isotropic and composite turbulent fluctuations (red lines). Also shown are some results of the simulations performed by \citet{Minnie_etal2007b} and \citet{ts2012} for composite turbulence conditions (data points).}
    \label{fig1}
\end{figure*}

In terms of the drift coefficient $\kappa_A$, Eq. \ref{aaap2} is therefore equal to
\begin{equation}
\langle \vec{v}_d \rangle = \nabla \times  \kappa_{A}^{ws} f_s \mathbf{e}_{B_0}   \label{aap3}
\end{equation}
with $\mathbf{e}_{B_0} := \vec{B_0}/B_0$ a unit vector along the mean uniform field $\vec{B_0}$, and $f_s$ some factor by which the weak-scattering value of the drift coefficient $\kappa_{A}^{ws}$ is altered. This leads us to conclude, from inspection of Eq. \ref{aaap2}, that the drift coefficient is suppressed by a factor given by
\begin{equation}
f_s := \frac{1}{1 + \langle   b^2\rangle /B_0^2  }. \label{aap4}
\end{equation}
Some care must be taken in the interpretation of $\langle   b^2\rangle$, as the exact nature of the implied time-averaging is not clear. One possible approach to this problem is as follows. Defining the total variance of the fluctuating field as
\begin{equation}
\delta B_{T}^{2} := \int_0^{\infty} g(\vec{k}) d\vec{k}, \label{aap5}
\end{equation}
where $g(\vec{k})$ denotes the turbulence power spectrum associated with the fluctuating magnetic field component, the drifting particle is only expected to be influenced by fluctuations on scales comparable to, or larger than, its Larmor radius. Hence, we define
\begin{equation}
\langle b^2\rangle := \int_0^{R_L^{-1}} g(\vec{k}) d\vec{k}. \label{aap5b}
\end{equation}
For ease of comparison between the results of this section and those of previous studies mentioned in the previous section, we introduce the factor
\begin{equation}
\epsilon := \frac{\langle b^2\rangle}{\delta B_{T}^{2}}
\end{equation}
where a comparison between Equations \ref{aap5} and \ref{aap5b} indicates that $\epsilon \leq 1$. Then  Eq. \ref{aap4} becomes
\begin{equation}
f_s = \frac{1}{1 + \epsilon \delta B_{T}^{2}/B_0^2  } \label{aap6}
\end{equation}
where $\delta B_{T}^{2}$ denotes the total variance as defined in Eq. \ref{aap5}. This result is similar to that of \citet{jokipii1993}. A brief consideration of various limits shows that Eq. \ref{aap6} satisfies, at least to first order, what is expected of such a reduction factor from prior simulations such as those performed by \citet{Minnie_etal2007b}. In the very low turbulence limit, where $\delta B_{T}^{2} \ll B_0^2$, we have that $f_s \approx 1$, which returns the weak scattering drift coefficient, while for the case where the turbulence is strong ($\delta B_{T}^{2} \gg B_0^2$), we have $f_s \rightarrow 0$. Furthermore, at low particle energies, $r_L^{-1}$ becomes very large, implying that $\epsilon$ approaches unity and hence that $f_s \rightarrow (1+\delta B_{T}^{2}/B_0^2)^{-1}$, so that there would be a maximum reduction of the weak scattering drift coefficient. Conversely, at high particle energies there is no drift reduction, as $r_L^{-1} \ll 1$, implying that $\epsilon \rightarrow 0$, which in turn yields $f_s \rightarrow 1$. Also, it is immediately apparent that the form of Eq. \ref{aap6} resembles strongly that of the functions \citet{ts2012} fit to their simulations of the turbulence-reduced drift coefficient. This is further reinforced by a cursory inspection of Fig. \ref{fig1}, which shows examples of $f_s$, as function of $\delta B_{T}^{2}/B_0^2$ for varying values of $\epsilon$, along with the simulation fits proposed by \citet{ts2012} for the cases of isotropic and composite ($85\% / 15\%$ 2D/slab) turbulence. Although the \citet{ts2012} fit for their reduction factor in the presence of isotropic turbulence falls below the $\epsilon=1$ case for Eq. \ref{aap6}, the composite result falls neatly within the range expected of that equation.

Also shown on the same figure are the results of numerical simulations performed by \citet{Minnie_etal2007b}, for two different ratios of the proton Larmor radius to the slab correlation scale assumed in that model such that $R_L/\lambda_c$ is equal to $0.1$ and $1.0$, as well as the results reported by \citet{ts2012} for $R_L/\lambda_{c}=0.1$. Note that these simulations were performed for the approximately the same composite turbulence conditions as those of \citet{Minnie_etal2007b}, the difference being that \citet{Minnie_etal2007b} assume $80\% / 20\%$ 2D/slab turbulence. It is clear that these simulation results fall within the range delineated by the limiting cases of $\epsilon=0$ and $1$.

The similarity of the drift reduction coefficient of Eq. \ref{aap6} in form to the fits presented by \citet{ts2012}, as well as the fact that the limiting cases for Eq. \ref{aap6} effectively bound the simulation results of that study as well as those of \citet{Minnie_etal2007b}, suggest that, at least to first order, the approach presented here will yield a reasonable approximation to the factor by which turbulence reduces the weak scattering drift coefficient, even though uncertainty implicit to the averaging performed on Eq. \ref{aap1} makes it difficult to accurately and self-consistently estimate the effect of turbulent fluctuations likely to affect the drift of the particles in question. Furthermore, the drift reduction coefficient of Eq. \ref{aap6}, which was derived without making assumptions as to the geometry of the turbulence apart from it being transverse to the background field, cannot explain the simulated drift coefficients reported by \citet{ts2012} for purely slab turbulence, which essentially remained at the weak scattering level, except by assuming \textit{a posteriori} that only 2D turbulent fluctuations act so as to reduce the drift coefficient. Lastly, the sensitivity of numerically simulated cosmic ray intensities demonstrated by \citet{EB2015} to the form of the turbulence-reduced drift coefficient employed also implies that a first-order result for $f_s$ may prove to be of limited use in modulation studies, given the uncertainty in the averaging of Eq. \ref{aap1}. The following section outlines an alternative approach to the calculation of this quantity, based on the work of \citet{bm1997}, which does not suffer from these limitations.

\section{A modification to the results of Bieber \& Matthaeus (1997)}
\label{sec-bm}

In their approach, \citet{bm1997} invoke the TGK (\citet{taylor}-\citet{green}-\citet{kubo}) formula for a diffusion coefficient in terms of some relevant velocity correlation function:

\begin{equation}
D_{ij}=\int_{0}^{\infty}dtR_{ij}(t)  \label{aap7}
\end{equation}

where the subscripts $i$ and $j$ denote Cartesian coordinates (in this study the background magnetic field is assumed to be uniform and pointed in the $z$-direction), and $R_{ij}(t)=\langle v_{i}(t_o)v_{j}(t_o+t)\rangle$ the velocity correlation function, which is assumed to be independent of the reference time $t_o$, and to go to zero at a rate greater than $1/t$ as $t$ goes to infinity. The assumption that the decay of this correlation function is a function of the time interval $(t+t_o)$ alone implies the assumption that particles are interacting with stationary, homogenous turbulence. \citet{bm1997} note that the calculation of this correlation function from first principles, that is to say without making the simplifying assumptions outlined in Section \ref{sec-2}, is extraordinarily difficult, as information is required as to the spatial and temporal dependences of the turbulent fluctuations. These authors proceed in their derivation of a turbulence-reduced drift coefficient by choosing physically and theoretically motivated forms for the required correlation functions, by considering the effect of magnetic fluctuations on the unperturbed gyromotion of a particle in a uniform magnetic field, arguing that such fluctuations would cause $R_{ij}$ to go to zero after a sufficient amount of time has elapsed. The form \citet{bm1997} choose of interest to this study is then
\begin{equation}
R_{yx}=\frac{v^2}{3}\sin(\Omega t)e^{-\nu_{\perp}t}  \label{aap8}
\end{equation}
with $\Omega$ the gyrofrequency of the unperturbed particle, $v$ its speed, and $\nu_{\perp}$ some perpendicular decorrelation rate. Integration of this correlation function in Eq. \ref{aap7} with $\nu_{\perp}=0$ then yields the weak-scattering drift coefficient, while for non-zero values of the decorrelation rate, it yields Eq. \ref{aap0} with $\tau=1/\nu_{\perp}$. \citet{bm1997} then argue that the field line random walk process will be the major factor in the perpendicular decorrelation process, introducing a lengthscale $z_c=R_{L}^{2}/D_{\perp}$ over which the perpendicular correlation function would significantly decrease. This then leads to a decorrelation time of
\begin{equation}
\tau \sim \frac{R_{L}^{2}}{vD_{\perp}}.  \label{aap9}
\end{equation}
This scaling forms the basis of the drift-reduction term proposed by these authors, as discussed in Section \ref{sec-1}. In the present study, we do not assume that decorrelation is entirely due to FLRW, as the drift process would act so as to cause particles to leave field lines. We assume that the perpendicular decorrelation scale is inversely proportional to some lengthscale along which decorrelation perpendicular to the uniform background field occurs, which we approximate as the particle's perpendicular mean free path, so that $z_c=R_{L}^{2}/\lambda_{\perp}$. The choice of $\lambda_{\perp}$, as opposed to the turbulence correlation length, is motivated by the fact that we are interested in the particle velocity decorrelation in particular. Furthermore, due to the fact that particles drift perpendicular to the background field, we assume that the perpendicular decorrelation rate is influenced only by the particle's speed perpendicular to the uniform background field $v_{\perp}$. This then gives the decorrelation time as
\begin{equation}
\tau = \frac{R_{L}^{2}}{v_{\perp}\lambda_{\perp}}.  \label{aap10}
\end{equation}
The perpendicular decorrelation speed is unaffected by the drift velocity term, as it will not contribute to this perpendicular speed under the assumption of a uniform constant background magnetic field, even in the presence of turbulent fluctuations, as indicated by the simulation results of \citet{Minnie_etal2007b}. To get an estimate of this perpendicular speed then, consider a Reynold's decomposed turbulent magnetic field in two dimensions $\vec{B}=B_{0}\vec{e}_{z}+b_{x}\vec{e}_{x}$, where $B_0$ is uniform, $b_x$ a fluctuating, transverse component, and $\langle B \rangle = B_0$. Then at any particular point along $\vec{B}$, the sine of the angle $\theta$ between $\vec{B}$ and $B_{0}\vec{e}_{z}$ will be given by $b_{x}/B \approx b_{x}/B_{0}$, assuming small fluctuations. This angle will be the same then as the average angle between the particle velocity $\vec{v}$ and it's component parallel to $\vec{e}_{z}$, such that $\sin \theta = v_{x}/v$, again assuming small fluctuations. This then leads to $v_{x} \approx v(b_{x}/B_{0})$. As it follows that $\langle v_{x} \rangle =0$, we then model $v_{\perp}$ as the root-mean-square value of this quantity. Therefore, we use $v_{\perp} \approx v(\delta B_{T}/B_{0})$, which then leads to
\begin{equation}
\Omega\tau = \frac{R_{L}}{\lambda_{\perp}}\frac{B_0}{\delta B_{T}},  \label{aap11}
\end{equation}
which, after substitution into Eq. \ref{aap0} and a little rearrangement, yields
\begin{equation}
f_s = \frac{1}{1 + \frac{\lambda_{\perp}^{2}}{R_{L}^{2}}\frac{\delta B_{T}^2}{B_0^2}}. \label{aap12}
\end{equation}
This expression is reminiscent of the form of the reduction term derived in Section \ref{sec-2}. Perpendicular particle transport has been shown from simulations \citet{qin2002a,qin2002b} to be subdiffusive in the presence of pure slab turbulence. In this case, then, the perpendicular diffusion coefficient, and thus the perpendicular mean free path, would be zero (see, e.g., \citet{Shalchi2006a}). It should be noted here that both the theoretical treatments of the drift coefficient in the presence of turbulence proposed by \citet{Stawicki2005} and \citet{leroux2007} predict that there will be no drift reduction in the presence of pure magnetostatic slab turbulence. In these conditions, then, Eq. \ref{aap12} automatically yields the weak-scattering result, as seen in the simulations of \citet{ts2012}, as $\lambda_{\perp}$ would be zero under these conditions \citep{qin2002a,Shalchi2006a}. The fact that Eq. \ref{aap12} is a function of the perpendicular mean free path, and thus implicitly of the parallel mean free path (assuming a nonlinear guiding center theory prediction for $\lambda_{\perp}$) is also in line with the findings of \citet{Minnie_etal2007b}, who report that knowledge of the spatial variation of these mean free paths would be required to fully describe particle drifts.

The asymptotic behaviour of this drift-reduction term now depends on the various implicit dependences of the perpendicular mean free path on, for example, the Larmor radius and the ratio of the variance to the background field strength. Assuming that $\lambda_{\perp}$ remains relatively uniform as function of rigidity (and therefore of $R_L$), as implied by both the \citet{palmer1982} consensus range as well as various numerical simulations (e.g. \citet{Minnieetal2007a}) and theoretical results (see, e.g., \citet{shalchibook}), the ratio $\lambda_{\perp}/R_L$ would correspond to small values of the quantity $\epsilon$ in Eq. \ref{aap6} at large energies, and large values of $\epsilon$ at the lowest energies, based on an assumed value of $\delta B_{T}^{2}/B_0^2$. This then would imply significant reduction of the drift coefficient from the weak-scattering value at low energies, and limited reduction at high energies, as expected from simulations. Furthermore, if one were to hold the ratio $\lambda_{\perp}/R_L$ constant, it is clear that the drift coefficient would be more reduced at high turbulence levels, and less reduced at the lowest values of $\delta B_{T}^{2}/B_0^2$, again as expected from simulations. We also do not expect a strong dependence of this drift-reduction factor on the spectral index of the energy-containing range of the 2D turbulence power spectrum, as from theoretical results (see, e.g., \citet{shalchietal2010} and \citet{EB2015b}) the rigidity dependence of $\lambda_{\perp}$ for different values of this spectral index is never as steep as that of $R_{L}^2$, again in qualitative agreement with the simulation results of \citet{ts2012}.  

\begin{figure*}
    \centering
    \includegraphics[width=0.95\textwidth]{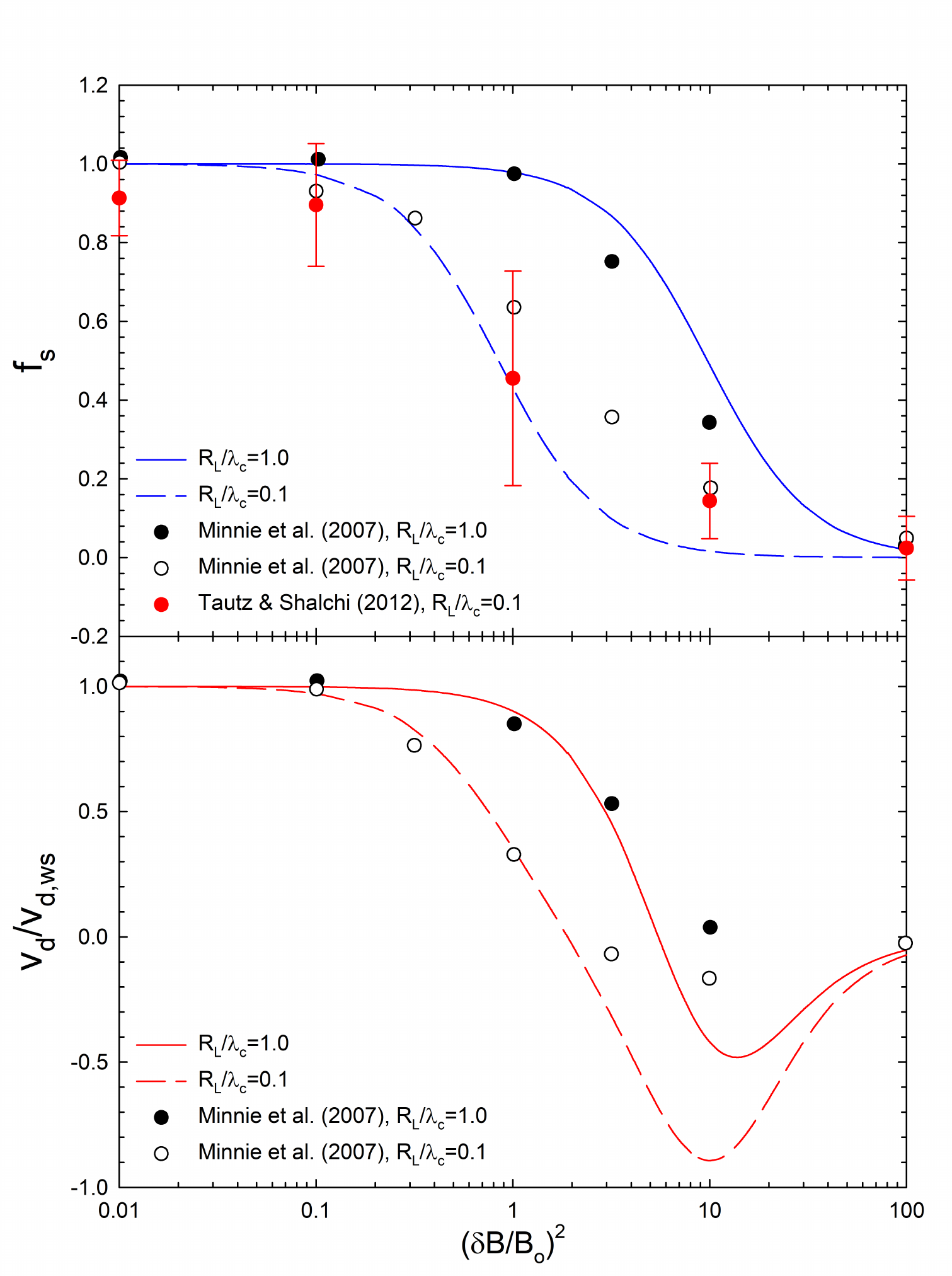}
    \caption{Drift reduction function of Eq. \ref{aap12} (top panel) and normalized drift speed (bottom panel) for different choices of $R_L/\lambda_c$ (blue lines). Also shown are some results of the simulations performed by \citet{Minnie_etal2007b} and \citet{ts2012} for composite turbulence conditions (data points).}
    \label{fig2}
\end{figure*}
It remains, however, to be seen whether Eq. \ref{aap12} can yield results comparable to those yielded by numerical simulations of the drift reduction term. In order to make this comparison, a choice needs to be made as to an expression for $\lambda_{\perp}$. This is not a trivial matter, as the implicit dependence of $\lambda_{\perp}$ on, e.g., turbulence quantities will have a significant effect on $f_s$. We choose an analytical approximation for the perpendicular mean free path derived from the nonlinear guiding center (NLGC) theory of \citet{Matthaeusetal2003} by \citet{shalchietal2004}, as modified by \citet{Burger2008}. This choice is motivated by the tractability of the expression, which allows for ease of comparison with simulation results (as opposed to the nonlinear results of, say, \citet{EB2013}), as well as being derived for a 2D turbulence spectral form identical to that employed as an input to the numerical simulations of \citet{Minnie_etal2007b} (and some of the simulations of \citet{ts2012}), which contains a flat energy-containing range, and a Kolmogorov inertial range. Furthermore, this result also automatically satisfies the Shalchi slab hypothesis (see \citet{Shalchi2006a}), as it becomes zero when the 2D variance is zero. This perpendicular mean free path expression is
\begin{equation}
\lambda_{\perp}=\left[\alpha^{2}\sqrt{3\pi}\frac{2\nu-1}{\nu}\frac{\Gamma (\nu)}{\Gamma (\nu-1/2)}\lambda_{2D}\frac{\delta B_{2D}^2}{B_{0}^2}\right]^{2/3}\lambda_{\parallel}^{1/3}  \label{aap13}
\end{equation}
where we assume that $\alpha^{2}=1/3$ (from \citet{Matthaeusetal2003}), $\lambda_{2D}$ is the turnover scale where the inertial range commences on the assumed 2D turbulence power spectrum, and $\nu$ denotes half the assumed inertial range spectral index. As input for the parallel mean free path we use a quasilinear theory expression based on the results of \citet{ts2003}:
 \begin{equation}
\lambda_{\parallel}=\frac{3s}{\pi (s-1)}\lambda_{s}R^{2}\frac{B^{2}_{o}}{\delta B^{2}_{sl}}\left[\frac{1}{4}+\frac{2R^{-s}}{(2-s)(4-s)}\right],
\label{eq:lamparprs}
\end{equation}
where $R=R_{L}/ \lambda_{s}$ is a function of the lengthscale at which the inertial range on the slab turbulence power spectrum commences, which is assumed to have a spectral index $s$. This choice is also motivated by the tractability of Eq. \ref{eq:lamparprs}, as well as the fact that it is derived assuming a slab spectral form similar to that assumed in the simulation results we are comparing our results to. In order to properly compare our result with the simulations of \citet{Minnie_etal2007b}, we choose values for turbulence parameters identical to those used in that study, so that $s= 2\nu =5/3$, $\delta B_{2D}^2=0.8\delta B_{T}^2$, $\delta B_{s}^2=0.2\delta B_{T}^2$ and $\lambda_{s}=10 \lambda_{2D} = 1.0$. The results of these choices for the parallel and perpendicular mean free paths as inputs for Eq. \ref{aap12}, using the values for the turbulence quantities listed above, are plotted as function of the ratio of $R_L$ to the slab correlation length $\lambda_c$ in the top panel of Fig. \ref{fig2}, along with the numerical simulation results of \citet{Minnie_etal2007b} and \citet{ts2012}, as function of the level of turbulence $\delta B_{T}^2/B_0^2$. Note that the slab correlation length is related, for the particular slab turbulence spectral form employed here, by $\lambda_c=\sqrt{\pi}\Gamma (\nu-0.5)\lambda_{s}/\Gamma (\nu)$. As expected, Eq. \ref{aap12} predicts that at higher particle energies, only the highest levels of turbulence cause a reduction in the drift coefficient. Agreement with the \citet{Minnie_etal2007b} simulations at $R_L/\lambda_{c}=1.0$ is good, but less so for $R_L/\lambda_{c}=0.1$. The latter prediction, however, falls within the error bars reported by \citet{ts2012}. Note that for lower levels of turbulence ($\delta B_{T}^2/B_0^2\lesssim 0.1$ and even to a lesser degree, given the extent of the uncertainties in the simulations, $\delta B_{T}^2/B_0^2 \lesssim 1$), Eq. \ref{aap12} is in good agreement with the simulation results. From observations \cite[e.g.][]{bieberetal1993,zanketal1996,Burger2014} and turbulence transport modelling \cite[e.g.][]{zank,tobias,usmanov} it is this range of turbulence levels that is typical in the heliosphere. It should be noted that, for their simulations assuming a gradient in the background magnetic field, \citet{Minnie_etal2007b} report turbulence-reduced drift coefficients essentially as those they calculate using a uniform background field. A comparison between the drift velocity calculated using Eq. \ref{aap12} and the simulation results of \citet{Minnie_etal2007b}, assuming a background magnetic field with a gradient, is shown in the bottom panel of Fig. \ref{fig2}. Here the drift velocities are calculated using $\vec{v}_{d}=\nabla \times \kappa_A \vec{e}_{B}=f_s \nabla \times \kappa^{ws}_{A} \vec{e}_{B}+ \nabla f_s \times \kappa^{ws}_{A} \vec{e}_{B}$, again assuming parameters identical to those employed by \citet{Minnie_etal2007b}, and normalised to the zero-turbulence drift velocity. Note that we only compare our results with those pertaining to the y-component of the drift velocities calculated by \citet{Minnie_etal2007b}, as the gradient imposed by these authors on their simulated background field (which points in the z-direction) has only an x-component. Here, the use of Eq. \ref{aap12} again leads to good agreement with simulation results at smaller levels of turbulence relevant to heliospheric conditions for both values of $R_L/\lambda_{c}$ considered. At higher turbulence levels, the y-component of the drift velocity calculated using Eq. \ref{aap12} does not agree well with the simulations, a consequence of the assumption of relatively weak turbulence in the derivation of that expression. It is interesting to note that the simulations for the case where $R_L/\lambda_{c}=0.1$ yield negative values for the y-component of the drift velocity, as is the case for the results calculated using Eq. \ref{aap12}, even though the latter approach overestimates this effect. However, given the range of turbulence levels relevant to the heliosphere as discussed above, such an effect would not be expected to have significant consequences as to the transport of charged particles.
\section{Discussion}

The form of the drift-reduction factor contained in Eq. \ref{aap12} provides a relatively simple, tractable way of describing and modelling the effects of a range of turbulence conditions on the drift coefficient of charged particles that satisfies the conditions prescribed by extant numerical simulations of both the drift coefficient and velocity as well as yielding results in reasonably good agreement with said simulation results for turbulence levels corresponding to what is expected in the heliosphere. Due to its explicit dependence on $\lambda_{\perp}$, this quantity should, if used in conjunction with an expression for the perpendicular mean free path and a turbulence transport model, yield complicated spatial dependences for $f_s$ throughout the heliosphere, as has been shown by \citet{tobias} for the drift-reduction factors discussed in Section \ref{sec-1}. The dependence of Eq. \ref{aap12} on basic turbulence quantities should have consequences for studies of the transport of particles such as low-energy electrons of galactic and Jovian origin. These particle's parallel and perpendicular mean free paths are expected to remain at  a relatively constant value for a given (small) rigidity \citep[see, e.g.][]{EB2013b}, which, in combination with the explicit Larmor radius dependence of Eq. \ref{aap12}, would lead to small values of $f_s$ for a given turbulence level, and thus lead to a greatly reduced drift coefficient relative to the diffusion coefficients in line with what is expected from prior modulation studies \citep[e.g][]{Potgieter1996}. Furthermore, the transport of solar energetic particles would also be affected, in that the higher levels of turbulence closer to the sun \citep[see, e.g.][]{Bruno2013} would feed into Eq. \ref{aap12} in such a way so as to kill off any drifts such particles may encounter, a prediction in contrast to the simulation results reported by, e.g, \citet{dalla2013} and \citet{dalla2015}. Lastly, the implicit dependence of Eq. \ref{aap12} on basic turbulence quantities leads to an implicit solar-cycle dependence for this drift-reduction factor. \citet{Burger2014} report an increase in the total magnetic variance at Earth as solar activity increases \cite[see also][]{bieberetal1993}. This increase would act so as to decrease $f_s$, and thus lead to greatly reduced drift effects during solar maximum as opposed to solar minimum, as expected from the modulation studies of, e.g., \citet{Ndii} and \citet{rex}.

Some caution has to be exercised in the use of Eq. \ref{aap12} due to the assumptions made as to the forms used for the perpendicular decorrelation lengthscale and speed that enter into Eq. \ref{aap10}. Furthermore, use of Eq. \ref{aap12} in modulation studies requires the assumption of some form for the perpendicular mean free path, which, given the number of expressions for this quantity currently in the literature \citep[see, e.g.,][]{shalchibook,ruffolo12,qz14}, can also lead to further uncertainty. To model the drift-reduction factor throughout the heliosphere would also require one to employ a turbulence transport model to provide information as to how the basic turbulence quantities $\lambda_{\perp}$ is a function of vary throughout the heliosphere. Lastly, Eq. \ref{aap12} does not take into account the possibility of non-axisymmetric turbulence, which could potentially play a role in the drift of charged particles \citep{Weinhorst2008}. These considerations point to the fact that the predictions of Eq. \ref{aap12} should be further tested, firstly by means of numerical test particle simulations, using as input for $\lambda_{\perp}$ the perpendicular mean free path calculated from the simulations themselves and assuming a broader range of turbulence conditions than that hitherto considered, and secondly, by means of particle transport studies such as the numerical study of cosmic ray modulation or solar energetic particle transport.

\acknowledgments

NEE, RDS and RAB acknowledge support from the National Research Foundation (Grant 96478). Opinions expressed and conclusions arrived at are those of the authors and are not necessarily to be attributed to the NRF.




\clearpage

\end{document}